\documentclass[letterpaper,11pt]{article}
\pdfoutput=1
\usepackage{jheppub}
\usepackage{multirow}
\usepackage{shuffle}

\usepackage{breqn}

\usepackage{caption}
\usepackage{subcaption}

\usepackage{xspace}
\usepackage[countmax]{subfloat}
\usepackage{enumitem}
\usepackage{amssymb}
\usepackage{amsmath}
\usepackage{cancel}
\usepackage{color}
\usepackage{braket}
\usepackage{graphicx}
\usepackage{multirow}
\usepackage{verbatim}
\usepackage{amsthm}
\usepackage{slashed}
\usepackage{wasysym}
\usepackage{simplewick}
\usepackage{mathtools}
\usepackage{soul}
\usepackage{xspace}

\usepackage{mathrsfs}

\def\beq{\begin{equation}}
\def\eeq{\end{equation}}
\def\bea{\begin{eqnarray}}
\def\eea{\end{eqnarray}}

\def\cM{\mathcal{M}}

\def\cO{\mathcal{O}}

\def\nn{{\nonumber}}

\newcommand{\Eq}[1]{Equation~\eqref{#1}}
\definecolor{darkgreen}{rgb}{0.13,0.55,0.13}

\DeclareRobustCommand{\Sec}[1]{Sec.~\ref{#1}}

\DeclareRobustCommand{\App}[1]{App.~\ref{#1}}

\DeclareRobustCommand{\Fig}[1]{Fig.~\ref{#1}}

\DeclareRobustCommand{\Eq}[1]{Eq.~(\ref{#1})}

\def\be{\begin{equation}}
\def\ee{\end{equation}}

  \newcount\hour \newcount\minute
  \hour=\time \divide \hour by 60 \minute=\time
  \newcommand{\todaytime}{\today \ -- \number\hour :\ifnum \minute<10 0\fi\number\minute}

\def\spa#1.#2{\left\langle#1\,#2\right\rangle}
\def\spb#1.#2{\left[#1\,#2\right]}

\def\feynsl#1{
  \setbox0=\hbox{/} \setbox1=\hbox{$#1$}
  \dimen0=\wd0 \advance\dimen0 by -\wd1 \divide\dimen0 by 2
  \ifdim\wd0>\wd1 \raise.15ex\copy0\kern-\wd0\kern\dimen0\copy1\kern\dimen0
  \else \kern-\dimen0\raise.15ex\copy0\kern-\dimen0\kern-\wd1\copy1\fi}

\newskip\humongous \humongous=0pt plus 1000pt minus 100pt

\newif\ifdtup

\def\beq{\begin{equation}}
\def\eeq{\end{equation}}
\def\beeq{\begin{eqnarray}}
\def\eeeq{\end{eqnarray}}

\def\Je2e{J_{E^2 E}}

\usepackage{xcolor}

\preprint{MIT-CTP-5605}

\title{The 1 $\rightarrow$ 3 Massive Splitting Functions from QCD Factorization and SCET}

\author[1]{Evan Craft,}
\author[1]{Mark Gonzalez,}
\author[2]{Kyle Lee,}
\author[1]{Bianka Me\c caj,}
\author[1]{Ian Moult}

\affiliation[1]{Department of Physics, Yale University, New Haven, CT 06511, USA\vspace{0.5ex}}
\affiliation[2]{Center for Theoretical Physics, Massachusetts Institute of Technology, Cambridge, MA 02139}

\emailAdd{evan.craft@yale.edu}
\emailAdd{mark.gonzalez@yale.edu}
\emailAdd{kylel@mit.edu}
\emailAdd{bianka.mecaj@yale.edu}
\emailAdd{ian.moult@yale.edu}

\abstract{Splitting functions are universal functions describing the collinear dynamics of gauge theories, and as such are crucial ingredients for a wide variety of calculations in perturbative QCD.
We present analytic results for the triple collinear splitting functions in QCD with a single massive parton. 
We derive the splitting functions using two distinct methods; first by expanding the squared matrix elements in the collinear limit, and secondly by using soft-collinear effective theory with massive quarks. We find agreement between these two approaches, providing a strong check of our results. 
Additionally, we also check all iterated and soft limits of our results, finding agreement with predictions from factorization.
Our results provide an important ingredient for higher order perturbative calculations involving massive partons, and for the description of the collinear dynamics of heavy flavor jets.
}

\begin{document} 

\maketitle

\section{Introduction}\label{sec:intro}

Universal functions appearing in limits of gauge theory scattering amplitudes and cross sections are key ingredients in perturbative calculations. A particular example are splitting functions \cite{Altarelli:1977zs,Catani:1998nv,Campbell:1997hg,Badger:2004uk,Bern:2004cz,Badger:2015cxa,Czakon:2022fqi,Catani:2003vu,DelDuca:2019ggv,DelDuca:2020vst,Sborlini:2013jba,Sborlini:2014mpa,Sborlini:2014kla}, which describe the universal dynamics of partons in the collinear limit. These play a central role in the description of the collinear substructure of jets, as well as in fixed order subtraction schemes.  Due to their importance in understanding the universal infrared behavior of amplitudes~\cite{Kosower:1999rx,Bern:1999ry,Bern:1998sc,Catani:2011st}, significant effort has been applied to their perturbative calculation. Following the seminal calculation of the tree-level $1\to2$ splitting functions in \cite{Altarelli:1977zs}, the triple collinear splitting functions were computed at tree level in \cite{Catani:1998nv,Campbell:1997hg}, and the complete set of ingredients describing the collinear limit to next-to-next-to-leading order, namely the two-loop $1\to 2$ splitting functions \cite{Badger:2004uk,Bern:2004cz}, the one-loop $1\to 3$ splitting functions \cite{Badger:2015cxa,Czakon:2022fqi,Catani:2003vu} and the tree level $1\to 4$ splitting functions \cite{DelDuca:2019ggv,DelDuca:2020vst}, were all computed for massless QCD partons.

Despite this tremendous progress in the perturbative description of the collinear limits of amplitudes and cross sections involving massless partons, considerably less is known about collinear limits involving massive partons. Indeed, only the $1\to 2$ massive splitting functions appear in compact forms in the literature \cite{Catani:2000ef}.   However, accurate perturbative calculations involving heavy quarks are crucial for many phenomenological applications. For example, the description of the substructure of jets involving $b$ quarks is crucial for studies of the Higgs and top-quark, and heavy quark jets are also ideal probes of the Quark Gluon Plasma (QGP) \cite{Andronic:2015wma,Rapp:2018qla,Dong:2019byy,Apolinario:2022vzg}, since they cannot be thermally produced. Additionally, massive splitting functions are a key ingredient in universal subtraction schemes for fixed order perturbative calculations involving massive partons. Combined, these strongly motivate an improved description of collinear limits involving massive partons in perturbation theory.

In this paper we compute the $1\to3$ splitting functions involving a single massive parton in QCD, and present our results in a compact form amenable to phenomenological applications. We derive our results in two independent ways. First, we derive them by expanding the relevant squared QCD matrix elements in the collinear limits, and comparing with the predictions of QCD factorization. Second, we derive them using soft-collinear effective theory (SCET)  \cite{Bauer:2001ct,Bauer:2000yr,Bauer:2001yt} with massive partons known as SCET$_{\text{M}}$ \cite{Leibovich:2003jd}. We find exact agreement between the two approaches, providing strong support to the correctness of our result. Additionally, we consider all iterated and soft limits of our results, and compare them with predictions from factorization. 

A key application of our results is to the calculation of jet substructure observables involving heavy quarks. In a companion paper, we will present the results for the three-point energy correlator \cite{Chen:2019bpb,Komiske:2022enw,Chen:2022swd} on heavy quark jets, which has interesting applications for better understanding the QCD dynamics of heavy quarks in both vacuum \cite{Craft:2022kdo} and the QGP \cite{Andres:2022ovj,Andres:2023ymw,Andres:2023xwr,Yang:2023dwc,Barata:2023zqg}. Additionally, there has been significant recent progress in the incorporation of mass effects into parton shower Monte Carlo simulations \cite{Assi:2023rbu}, and it would be interesting to combine this with approaches to incorporate triple collinear splitting functions in parton showers. Our results are a key ingredient towards achieving this goal.

An outline of this paper is as follows. In \Sec{sec:factorization}, we briefly review collinear factorization in gauge theories, and the universality of the splitting function. In \Sec{sec:calculation}, we describe our calculation of the triple collinear splitting function, both from the collinear expansion of squared QCD matrix elements, as well as using SCET with masses. In \Sec{sec:results}, we  present our results for the $1 \rightarrow 3$ splitting functions involving a single massive parton. In \Sec{sec:iterated} we check the iterated and soft limits of our results against predictions from factorization. Additional perturbative ingredients are collected in the Appendices for completeness.

{\bf{Note Added:}} On the day of submission of this manuscript, \cite{Dhani:2023uxu} appeared, which also computed the triple collinear splitting functions involving a single massive parton in QCD by expanding the squared matrix element. In preliminary comparisons, we find agreement with their results. Additionally, our calculation using SCET provides strong evidence for the correctness of both results.

\section{Factorization for Massive Collinear Limits}\label{sec:factorization}

\begin{figure}
	\centering
	\includegraphics[width=0.65\textwidth]{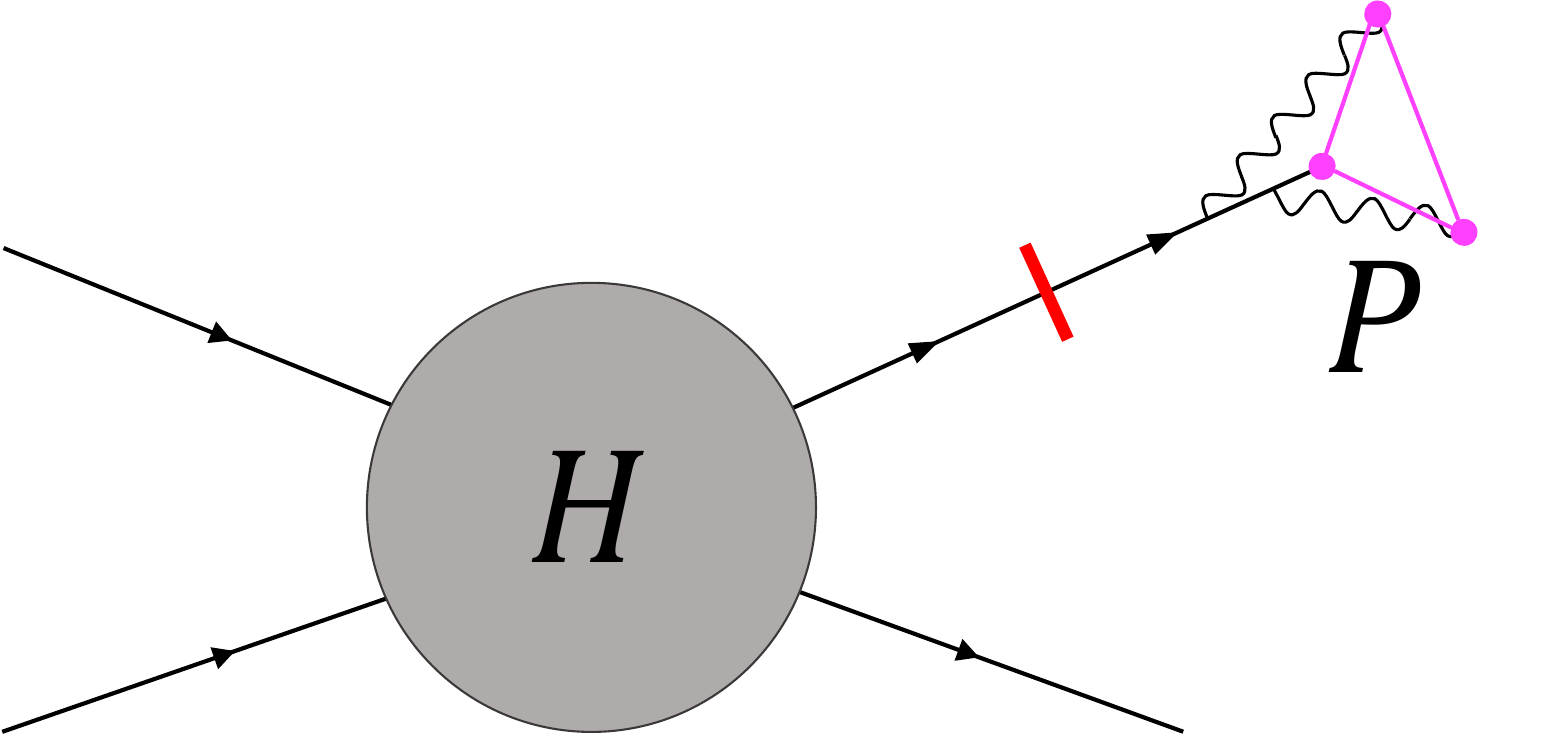}
	\caption{In the collinear limit, matrix elements factorize into a universal splitting function $P$ which describes the n-collinear partons, as well as a hard function $H$ describing the other non-collinear partons. The factorization on to on-shell states is represented by the red line.}
\end{figure}

It is well known that QCD amplitudes with a subset of collinear particles can be written in a factorized form (see, e.g., \cite{Ellis:1978sf,Amati:1978by,Amati:1978wx,Catani:1998nv}), where the collinear dynamics are encoded entirely in universal functions that depend only on the kinematics and quantum numbers of the collinear particles. Our focus in this work lies in the triple collinear limit of the matrix element, specifically when at least one of the collinear partons is a heavy quark $Q$. In this context, the squared matrix element is expressed as a product of a tree-level squared matrix element with $n-3$ external legs and a splitting kernel that describes the collinear splitting of a heavy quark into three partons $Q (p) \rightarrow a_1(p_1)+a_2(p_2)+a_3(p_3)$
\be\label{eq:massivefactorization}
\mathscr{C}_{123}\left|\mathcal{M}_{n}\left(p_1, \ldots p_n\right)\right|^2 =\left(\frac{2 \mu^{2 \epsilon} g_s^2}{\tilde{s}_{123}}\right)^{2} \left|\mathcal{M}_{n-3} (p_4,\ldots,p_n) \right|^{2} {P}_{a_1 a_2 Q}\left(x_i, p_{\perp}, m ; \epsilon\right)\,.
\ee   
Here, $\mathscr{C}_{123}$ represents the label~\cite{DelDuca:2019ggv} for the triple collinear limit, which we will describe in detail below, and ${P}_{a_1 a_2 Q}\left(x_i, p_{\perp}, m ; \epsilon\right)$ denotes the heavy quark triple collinear splitting kernels in $d=4-2\epsilon$ dimensions. When the parent parton is a quark, both the helicity and the fermion number must be conserved. This conservation implies the absence of non-trivial spin correlation between $P_{a_1 a_2 Q}$ and $\mathcal{M}_{n-3}$. Additionally, at least one of the outgoing partons must be a heavy quark $Q$, which we have labeled as the third parton without loss of generality.

The triple collinear splitting function carries spin, color, and flavor indices and is solely dependent on collinear kinematics. To derive this splitting function in the triple collinear limit $\mathscr{C}_{123}$, we introduce a light-cone vector $n^\mu$ that aligns with the parent parton's direction. We also introduce its conjugate vector $\bar{n}^\mu$ with the properties $n^2=\bar{n}^2=0$ and $n\cdot \bar{n}=2$, which then allows us to parameterize the collinear momenta as $p_i^\mu=\left(p_i^{+}, p_i^{-}, p_{\perp,i}\right)$ with $p_i^{+}=n \cdot p_i, p_i^{-}=\bar{n} \cdot p_i$. In other words
\begin{equation}\label{eq:collinear_momenta}
\begin{aligned}
	p_i^{\mu} &= \frac{\vec{p}_{\perp, i}^{\:2}}{x_i p^-} \frac{\bar{n}^{\mu}}{2} + x_i p^- \frac{n^{\mu}}{2} + p_{\perp, i}^{\mu}\,,  \qquad i=1,2\\
	p_3^{\mu} &=\frac{\vec{p}_{\perp, 3}^{\:2}+m^2}{x_3 p^-} \frac{\bar{n}^{\mu}}{2} + x_3 p^- \frac{n^{\mu}}{2} +  p_{\perp, 3}^{\mu}\,, 
\end{aligned}
\end{equation}
where $x_i$ are the momentum fractions of the on-shell light-like momenta $p^\mu = (q,0,0,q) = p^- n^\mu/2$ formed from the momentum direction of the parent parton $P^\mu=\sum p_i^\mu = (E-q,0,0,q)$. 

The triple collinear splitting kernels are then a function of the modified Mandelstam variables formed by these collinear momenta
\begin{equation}\label{eq:mandelstam}
\begin{aligned}
	\tilde{s}_{ijk} &\equiv s_{ijk} - m_i^2-m_j^2-m_k^2 \nonumber\\
	&= (p_i + p_j + p_k)^2 - m_i^2-m_j^2-m_k^2\,, \\
	\tilde{s}_{ij} &\equiv s_{ij} - m_i^2 - m_j^2 \nonumber\\
	&= (p_i+p_j)^2 - m_i^2 - m_j^2  \,.
\end{aligned}
\end{equation}
Here, we have $\tilde{s}_{12} = s_{12} = (p_1 + p_2)^2$, as the massive parton does not enter for $\tilde{s}_{12}$. Furthermore, $\tilde{s}_{123} = \tilde{s}_{12} + \tilde{s}_{13} + \tilde{s}_{23}$. In the following section, we compute the triple collinear splitting kernels at the leading order in $\alpha_s$.

\section{Description of the Calculation Methods}\label{sec:calculation}

\begin{figure}
	\centering
	\begin{subfigure}[b]{0.24\textwidth}
		\centering
		\includegraphics[width=\textwidth]{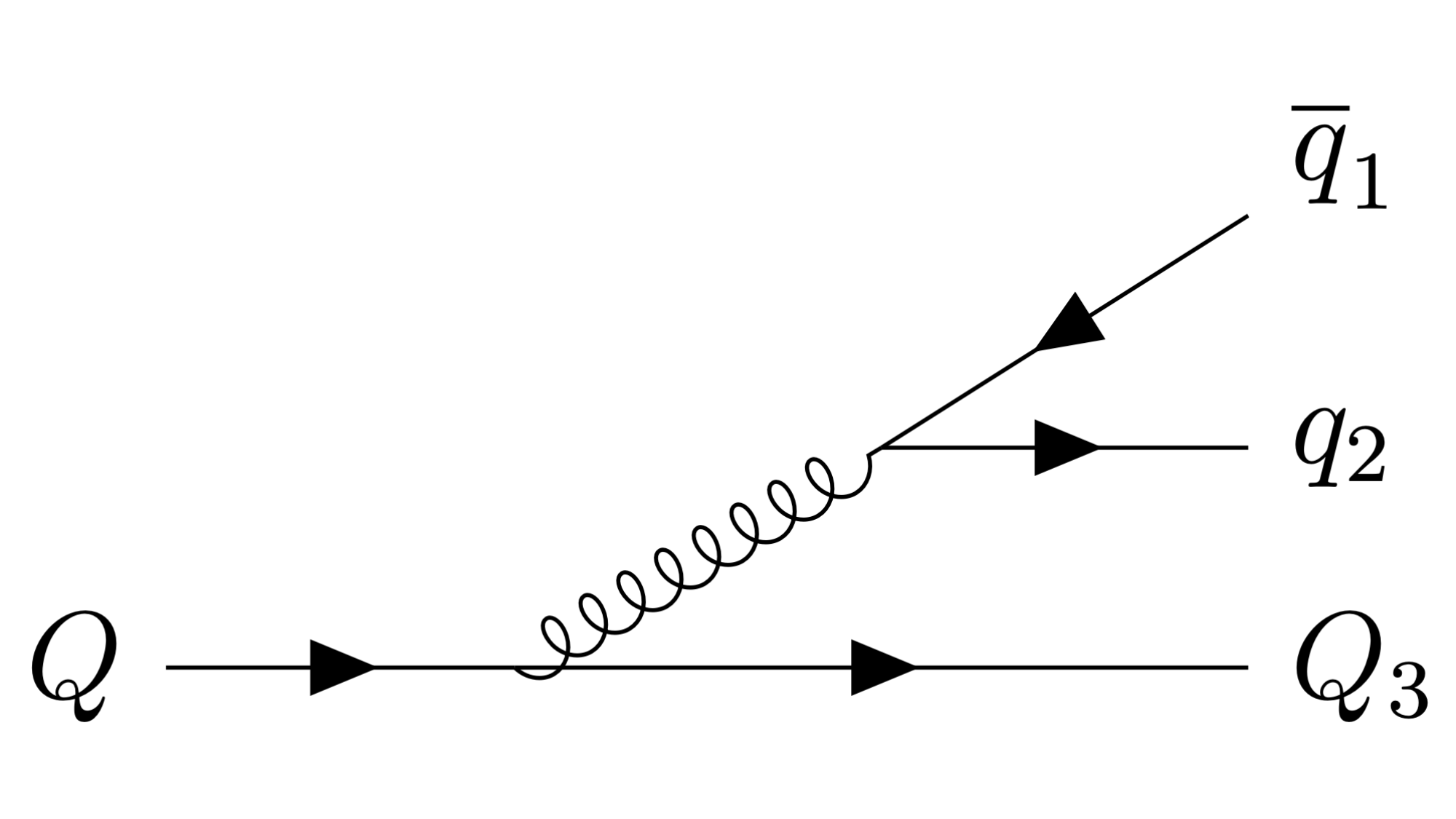}
		\caption{}
		\label{fig:FD_qqQ}
	\end{subfigure}
	\hfill
	\begin{subfigure}[b]{0.24\textwidth}
		\centering
		\includegraphics[width=\textwidth]{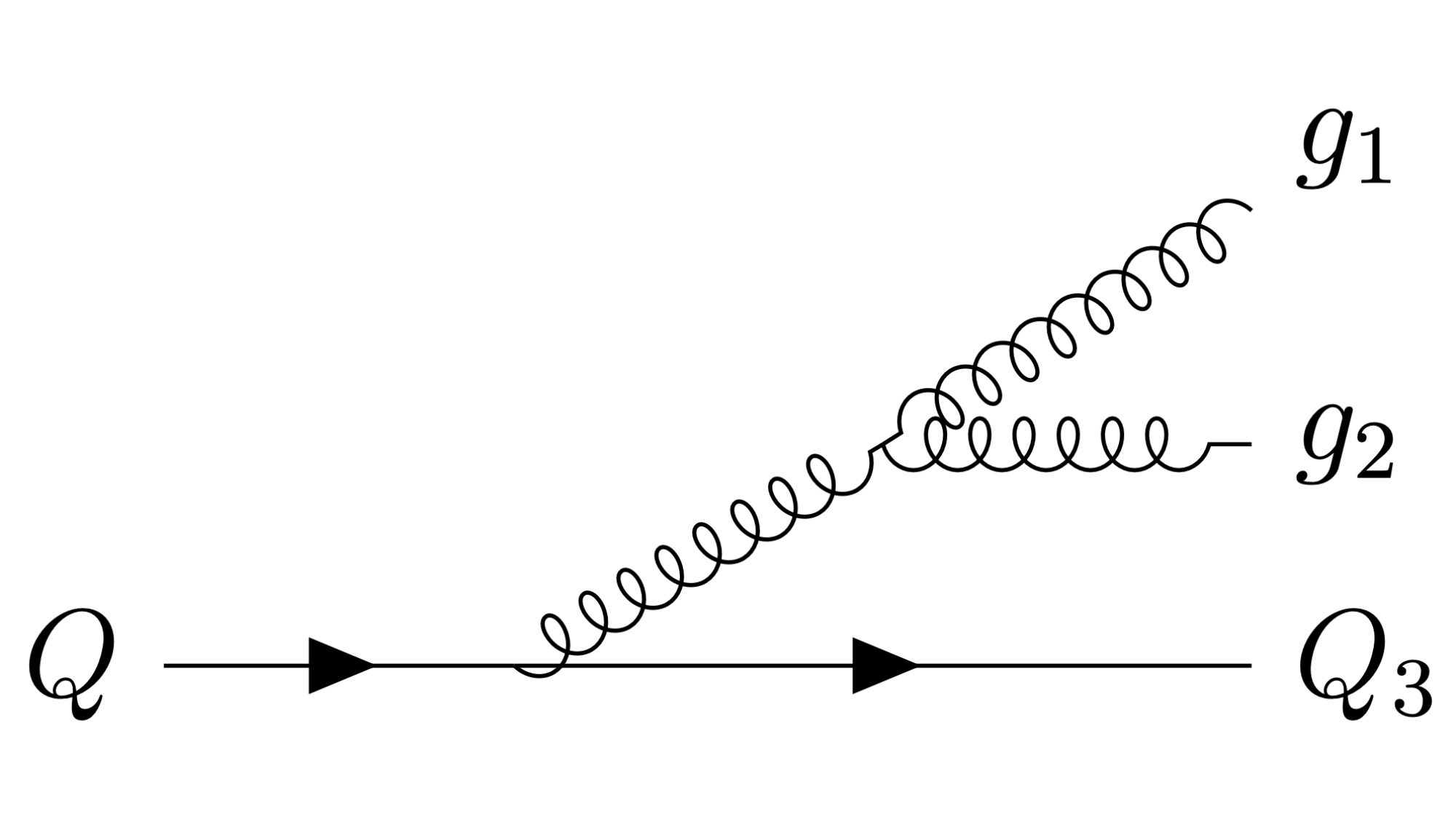}
		\caption{}
		\label{fig:FD_ggQNab}
	\end{subfigure}
	\hfill
	\begin{subfigure}[b]{0.24\textwidth}
		\centering
		\includegraphics[width=\textwidth]{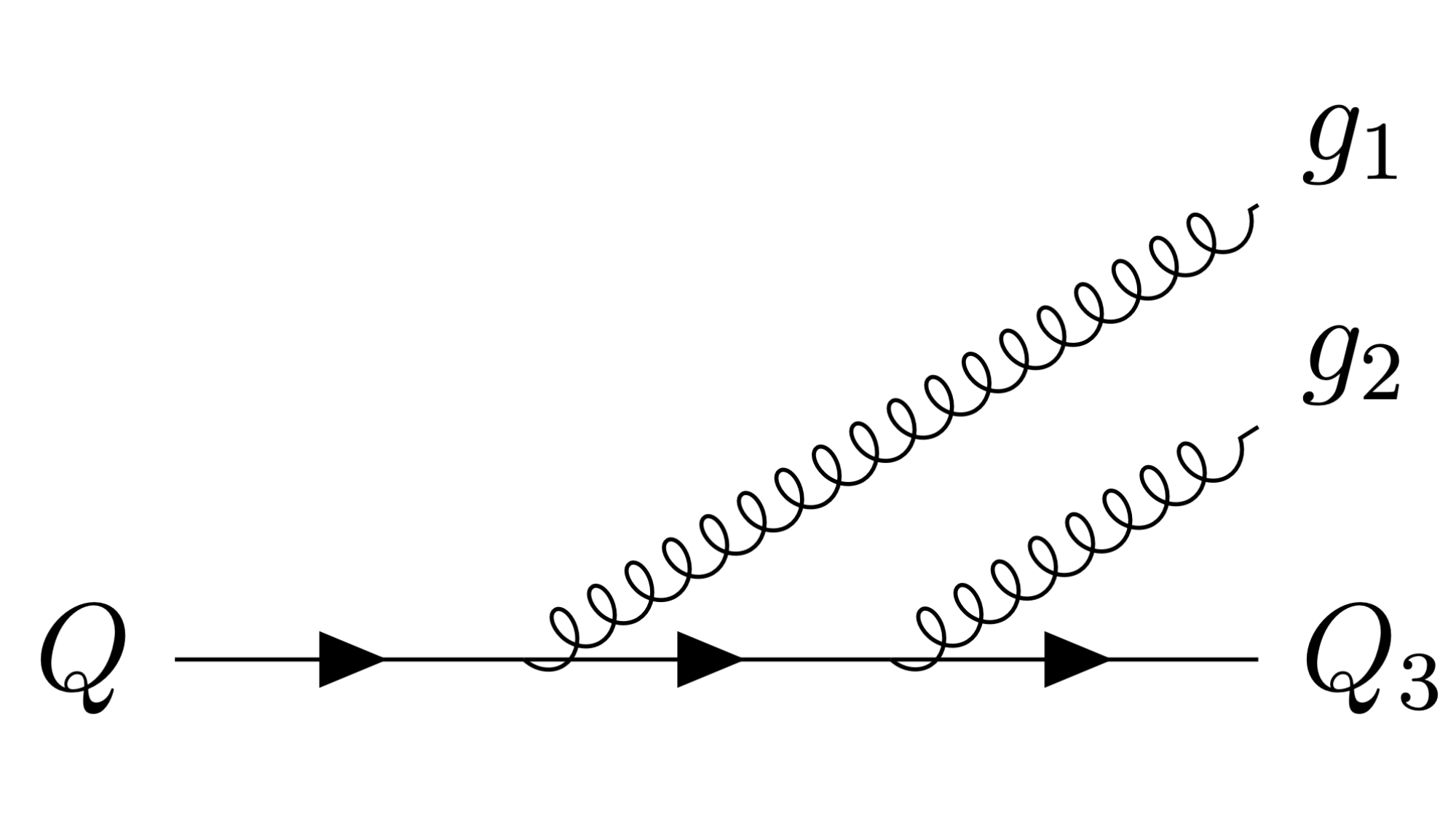}
		\caption{}
		\label{fig:FD_ggQ1}
	\end{subfigure}
	\hfill
	\begin{subfigure}[b]{0.24\textwidth}
		\centering
		\includegraphics[width=\textwidth]{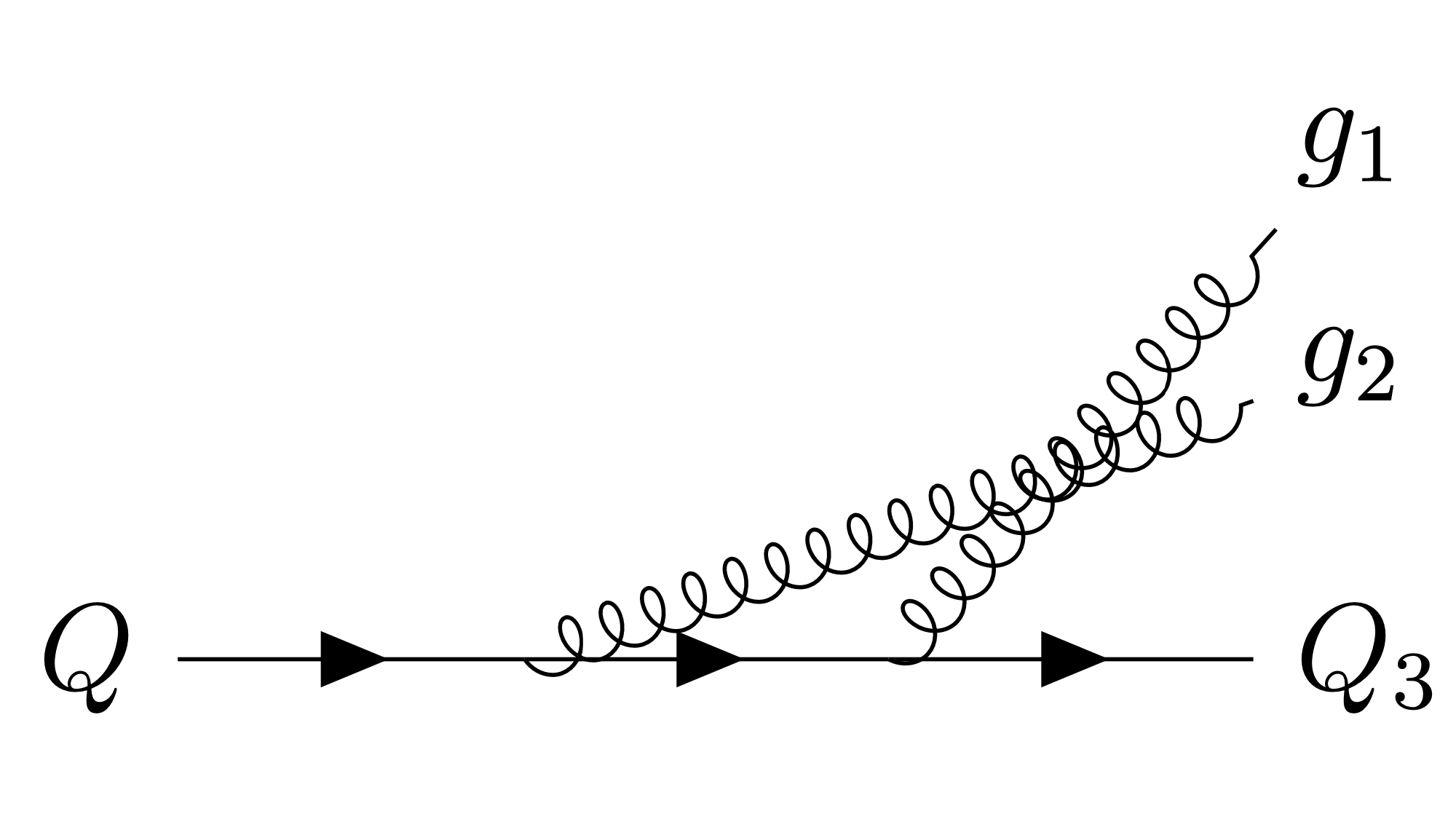}
		\caption{}
		\label{fig:FD_ggQ2}
	\end{subfigure}
	\caption{Feynman diagrams for various splitting channels in lightcone gauge. \Fig{fig:FD_qqQ} is the only diagram for $Q \rightarrow \bar{q}q Q$. \Fig{fig:FD_ggQNab}, \Fig{fig:FD_ggQ1}, and \Fig{fig:FD_ggQ2} are the diagrams for $Q \rightarrow ggQ$.}
\end{figure}

The universal collinear splitting functions given in~\Eq{eq:massivefactorization} can be obtained through several calculational methods. In Sec.~\ref{sec:expand_QCD}, we directly compute them by taking a collinear expansion of a relevant squared QCD matrix element and comparing it directly with the factorization formula given in \Eq{eq:massivefactorization}. In Sec.~\ref{sec:SCET}, we compute them by using SCET.


\subsection{Collinear Expansion of the Squared Matrix Elements}\label{sec:expand_QCD}

The splitting functions follow from taking the collinear limit of QCD amplitudes. The singular contributions in the collinear limit can be obtained by introducing an expansion parameter $\lambda$, and scaling the collinear momenta $p_j$ given in Eq.~\eqref{eq:collinear_momenta} by $(p_j^+,p_j^-,p_{\perp,j}) \sim p^-(\lambda^2,1, \lambda)$. Therefore, in the collinear limit, the massive parton has mass scaling like its transverse momenta, $m\sim p_{\perp,j} \sim \lambda\, p^- $. Given these scalings, we find that the modified Mandelstams scale as
\begin{equation}
	\tilde{s}_{123},\, \tilde{s}_{12},\, \tilde{s}_{13},\, \tilde{s}_{23} \sim \lambda^2 (p^{-})^2 \,.
\end{equation}
We then take the collinear limit $\mathcal{C}_{123}$ by deriving the singular limit as the expansion parameter $\lambda \to 0$. 
Scaling the quantities in the squared matrix elements as shown above, we find that the most singular terms go as
\begin{equation}
	|\cM|^2 \sim \cO(1/\lambda^4)\,.
\end{equation}
These singular terms then factorize as given in Eq.~\eqref{eq:collinear_momenta}, giving us the massive $1 \rightarrow 3$ splitting function.
The $Q \rightarrow Q q \bar{q}$ and $Q \rightarrow Qgg$ splitting functions were calculated with different gauge choices. Of course, we expect the collinear limit to be gauge invariant. The $Q \rightarrow Q q \bar{q}$ splitting function was calculated in Feynman gauge where the gluon propagator takes the simple form
\begin{equation}
    \frac{-i g^{\mu \nu} \delta^{a b}}{p^2 + i 0}\,.
\end{equation}
For the $Q \rightarrow Qgg$ calculation, we used the light-cone gauge as it reduces the number of diagrams that enter the calculation. In this case, the gluon propagator and polarization tensor, respectively, take the forms
\begin{equation}
    \frac{-i \delta^{a b}}{p^2 + i 0}\left(g^{\mu \nu} - \frac{\bar{n}^{\mu} p^{\nu} + \bar{n}^{\nu} p^{\mu}}{\bar{n} \cdot p}\right)\,,
\end{equation}
and
\begin{equation}
    \sum_{\text{polarizations}} (\epsilon^{\mu})^* \epsilon^{\nu} \equiv d^{\mu \nu}(p) =  -g^{\mu \nu} + \frac{\bar{n}^{\mu} p^{\nu} + \bar{n}^{\nu} p^{\mu}}{\bar{n} \cdot p}\,.
\end{equation}

\subsection{SCET with Masses}\label{sec:SCET}
The leading power SCET Lagrangian offers a more efficient method for calculating the leading triple collinear behavior. As the collinear expansion is already carried out at the Lagrangian level, once expressed in terms of the matrix elements of the gauge-invariant collinear SCET fields, there is no need to further carry out the collinear limit of the matrix elements. The $n$-collinear gauge-invariant building blocks for quark and gluon fields in SCET are given by~\cite{Bauer:2001ct,Bauer:2000yr,Bauer:2001yt}
\begin{align}
\chi_n=W_n^{\dagger} \xi_n, \qquad \mathcal{B}_{n \perp}^\mu=\frac{1}{g}\left[W_n^{\dagger} i D_{n \perp}^\mu W_n\right]\,,
\end{align}
where $ \xi_n$ and $A_{n}$ denote the $n$-collinear quarks and gluons fields, respectively, and $W_n$ denotes the Wilson line of collinear gluons. Here $i D_{n \perp}^\mu=\mathcal{P}_{n \perp}^\mu+g A_n^\mu$ and $\mathcal{P}$ is the label momentum operator in SCET. The Wilson line is given by 
\begin{align}
W_n(x)=\sum_{\text {perms }} \exp \left[-g \frac{1}{\bar{n} \cdot \mathcal{P}} \bar{n} \cdot A_n(x)\right]\,.
\end{align}
These gauge-invariant building blocks have a well defined scaling with respect to the parameter $\lambda$. This ensures that the expansion at the level of the fields in SCET is equivalent to the expansion in kinematic limits of QCD amplitudes. Using matrix elements formed from these gauge-invariant collinear building blocks of SCET fields at leading order in $\lambda$ we are therefore able to derive the triple collinear splitting functions directly from the SCET Lagrangian.

Some examples using this approach for the $1\to 2$ splitting functions can be found in \cite{Ovanesyan:2011kn,Kang:2016ofv}, and for a further discussion of the relation between splitting functions and jet functions in SCET, see \cite{Ritzmann:2014mka}. As a second approach to compute the triple collinear splitting functions, we computed the appropriate matrix elements of the gauge-invariant collinear fields from SCET with massive partons, known as SCET$_{\text{M}}$ \cite{Leibovich:2003jd}, to derive the massive $1\to 3$ splitting functions. In our calculation, we used the light-cone gauge throughout and found the expressions for the triple collinear splitting kernels computed using the SCET framework to agree with the results we obtained by taking the collinear expansion of the squared matrix elements using full QCD.

\section{Results for $1\to 3$ Massive Splitting Functions}\label{sec:results}

In this section we present our results for the $1\to 3$ splitting functions involving a single massive QCD parton. When expressed in terms of the variable $\tilde{s}$, the splitting functions involving a single massive parton can be organized into terms that are mass independent, or proportional to even powers of $m$. Since the mass independent splitting functions are familiar, we explicitly perform this separation to derive compact results. For each triple collinear splitting kernel, $P_{a_1a_2Q}$, we write the result as a sum
\begin{equation}
	P_{a_1a_2 Q} = R_{a_1a_2 Q} + M_{a_1a_2 Q}\,,
\end{equation}
where $R_{a_1a_2 Q}$ is identical to the known massless triple collinear splitting functions~\cite{Catani:1998nv} but with $s$ modified to $\tilde{s}$, whereas explicit mass dependent terms are captured by $M_{a_1a_2 Q}$. Since in the limit where $m \rightarrow 0$, we also have $\tilde{s} \rightarrow s$ and the $M$ term drops out, our splitting function agrees exactly with the massless splitting functions in the massless limit. Also, since the quark-initiated processes have trivial helicity dependence due to helicity conservation, our triple collinear splitting functions are always spin-averaged even without a $\langle~\rangle$ notation. In Sec.~\ref{sec:iterated} below, we compute the iterated limits with the polarization of the intermediate gluon intact. We now present the new $M$ terms as well as the known $R$ terms for completeness.

\subsection{$Q \rightarrow \bar{q}qQ$}

For the splitting process $Q \rightarrow \bar{q}qQ$, the mass independent term is given by \cite{Catani:1998nv} 
\begin{equation}
\begin{aligned}
	R _ {\bar{q} q Q} = C _ { F } T _ { F } \frac { \tilde{s} _ { 123 } } { 2 \tilde{s} _ { 12 } } \Bigg[ - \frac { \left[x_1 \left( \tilde{s} _ { 12 } + 2 \tilde{s} _ { 23 } \right) - x_2 \left( \tilde{s} _ { 12 } + 2 \tilde{s} _ { 13 } \right) \right] ^ { 2 } } { \left( x_1 + x_2 \right) ^ { 2 } \tilde{s} _ { 12 } \tilde{s} _ { 123 } } &+ \frac { 4 x_3 + \left( x_1 - x_2 \right) ^ { 2 } } { x_1 + x_2 } \nonumber\\&+ ( 1 - 2 \epsilon ) \left(x_1 + x_2 - \frac { \tilde{s} _ { 12 } } { \tilde{s} _ { 123 } } \right) \Bigg]\,,
\end{aligned}
\label{eq:qbqQml}
\end{equation}
while the mass dependent term, which is the novel result here, is simply given by
\begin{equation}
    M_{\bar{q}qQ} = - C_F T_R \frac{2 m^2}{\tilde{s}_{12}}\,.
\end{equation}

\subsection{$Q \rightarrow ggQ$}

As with its massless counterpart, the splitting function for the $Q \rightarrow ggQ$ process can be decomposed into its Abelian and non-Abelian color factors
\begin{equation}
    P_{ggQ} = C_F^2 P_{ggQ}^{(\text{ab})} + C_F C_A P_{ggQ}^{(\text{nab})}\,.
\end{equation}
The mass independent splitting function is then given by \cite{Catani:1998nv}
\begin{equation}
\begin{aligned}
	R _ {ggQ} =& C _ { F } ^ { 2 } \left\{ \frac { \tilde{s} _ { 123 } ^ { 2 } } { 2 \tilde{s} _ { 13 } \tilde{s} _ { 23 } } x_3 \left[ \frac { 1 + x_3 ^ { 2 } } { x_1 x_2 } - \epsilon \frac { x_1 ^ { 2 } + x_2 ^ { 2 } } { x_1 x_2 } - \epsilon ( 1 + \epsilon ) \right] + ( 1 - \epsilon ) \left[ \epsilon - ( 1 - \epsilon ) \frac { \tilde{s} _ { 23 } } { \tilde{s} _ { 13 } } \right] \right.\\&\left.+ \frac { \tilde{s} _ { 123 } } { \tilde{s} _ { 13 } } \left[ \frac { x_3 \left( 1 - x_1 \right) + \left( 1 - x_2 \right) ^ { 3 } } { x_1 x_2 } - \epsilon \left( x_1 ^ { 2 } + x_1 x_2 + x_2 ^ { 2 } \right) \frac { 1 - x_2 } { x_1 x_2 } + \epsilon ^ { 2 } \left( 1 + x_3 \right) \right] \right\}\\& + C _ { F } C _ { A } \left\{ ( 1 - \epsilon ) \left( \frac { \left[ x_1 \left( \tilde{s} _ { 12 } + 2 \tilde{s} _ { 23 } \right) - x_2 \left( \tilde{s} _ { 12 } + 2 \tilde{s} _ { 13 } \right) \right] ^ { 2 } } { 4 \left( x_1 + x_2 \right) ^ { 2 } \tilde{s} _ { 12 } ^ { 2 } } + \frac { 1 } { 4 } - \frac { \epsilon } { 2 } \right) \right.\\& + \frac { \tilde{s} _ { 123 } ^ { 2 } } { 2 \tilde{s} _ { 12 } \tilde{s} _ { 13 } } \left[ \frac { 2 x_3 + ( 1 - \epsilon ) \left( 1 - x_3 \right) ^ { 2 } } { x_2 } + \frac { 2 \left( 1 - x_2 \right) + ( 1 - \epsilon ) x_2 ^ { 2 } } { 1 - x_3 } \right]\\& - \frac { \tilde{s} _ { 123 } ^ { 2 } } { 4 \tilde{s} _ { 13 } \tilde{s} _ { 23 } } x_3 \left[ \frac { 2 x_3 + ( 1 - \epsilon ) \left( 1 - x_3 \right) ^ { 2 } } { x_1 x_2 } + \epsilon ( 1 - \epsilon ) \right]\\& + \frac { \tilde{s} _ { 123 } } { 2 \tilde{s} _ { 12 } } \left[ ( 1 - \epsilon ) \frac { x_1 \left( 2 - 2 x_1 + x_1 ^ { 2 } \right) - x_2 \left( 6 - 6 x_2 + x_2 ^ { 2 } \right) } { x_2 \left( 1 - x_3 \right) } + 2 \epsilon \frac { x_3 \left( x_1 - 2 x_2 \right) - x_2 } { x_2 \left( 1 - x_3 \right) } \right]\\&+ \frac { \tilde{s} _ { 123 } } { 2 \tilde{s} _ { 13 } } \left[ ( 1 - \epsilon ) \frac { \left( 1 - x_2 \right) ^ { 3 } + x_3 ^ { 2 } - x_2 } { x_2 \left( 1 - x_3 \right) } - \epsilon \left( \frac { 2 \left( 1 - x_2 \right) \left( x_2 - x_3 \right) } { x_2 \left( 1 - x_3 \right) } - x_1 + x_2 \right) \right.\\&\left.- \frac { x_3 \left( 1 - x_1 \right) + \left( 1 - x_2 \right) ^ { 3 } } { x_1 x_2 } + \epsilon \left( 1 - x_2 \right) \left.\left( \frac { x_1 ^ { 2 } + x_2 ^ { 2 } } { x_1 x_2 } - \epsilon \right) \right] \right\} + ( 1 \leftrightarrow 2 )\,.
\end{aligned}
\end{equation}
The mass dependent term is also symmetric under the exchange $1 \leftrightarrow 2$, however, we perform this symmetrization explicitly for clarity. We find
\begin{equation}
\begin{aligned}
    M_{ggQ}=& C_F^2 \Bigg\{\frac{4 m^4 (\tilde{s}_{13} + \tilde{s}_{23})^2}{\tilde{s}_{13}^2 \tilde{s}_{23}^2} - \frac{2 m^2 \tilde{s}_{123}}{\tilde{s}_{13}^2 \tilde{s}_{23}^2} \Bigg[(x_1 \tilde{s}_{13}^2+x_2 \tilde{s}_{23}^2)(1-\epsilon)-\epsilon(x_1+x_2) \tilde{s}_{13} \tilde{s}_{23}
     \\ &+ \frac{1}{x_1 x_2} \Big[2 x_1(x_1+x_3) \tilde{s}_{23}^2 + 2 x_2(x_2+x_3) \tilde{s}_{13}^2 + (x_1+x_2)(x_1 x_2+2 x_3)\tilde{s}_{13} \tilde{s}_{23}\Big] 
     \\ &- \frac{2 \tilde{s}_{13} \tilde{s}_{23}(\tilde{s}_{13}+\tilde{s}_{23})}{\tilde{s}_{123}}\Bigg]\Bigg\} \\
	&+ C_F C_A \Bigg\{ -\frac{4 m^4}{\tilde{s}_{13} \tilde{s}_{23}} - \frac{m^2}{x_1 x_2 (x_1+x_2) \tilde{s}_{12} \tilde{s}_{13} \tilde{s}_{23}}\Bigg[\tilde{s}_{12}^2 (x_1+x_2)^2 \Big[x_1 x_2 (\epsilon-1) - 2 x_3\Big] 
    \\& + 2 (\tilde{s}_{13}^2 x_2 + \tilde{s}_{23}^2 x_1)(x_1^2+x_2^2+x_1 x_2) + 2 \tilde{s}_{13} \tilde{s}_{23} (x_1^3+x_2^3) 
    \\&+ \tilde{s}_{12} \tilde{s}_{13} \Big[2 x_2^3 - x_2^2(1+\epsilon-2 \epsilon x_3-(1-\epsilon)x_3^2) \\&+ (x_1+x_2)^2 x_2(1+\epsilon+(1-\epsilon)x_3) - 2 (x_1+x_2)^2 x_3\Big] 
    \\&+ \tilde{s}_{12} \tilde{s}_{23} \Big[2 x_1^3 - x_1^2(1+\epsilon-2 \epsilon x_3-(1-\epsilon)x_3^2) \\&+ (x_1+x_2)^2 x_1(1+\epsilon+(1-\epsilon)x_3) - 2 (x_1+x_2)^2x_3\Big]\Bigg]\Bigg\}\,.
\end{aligned}
\end{equation}
The results in this section provide the complete set of $1\to 3$ splitting functions that involve a single massive parton necessary for phenomenological applications. In \App{app:B}, we also provide the $Q\to \bar{Q}'Q'Q$ and $Q\to \bar{Q}QQ$ splitting functions computed using our calculation methods. In all cases, we find that the explicitly mass dependent terms given in $M$ are of a similar level of complexity as the mass-independent terms given in $R$.
\section{Decomposition of $1\to 3$ Massive Splitting Functions}\label{sec:iterated}

In addition to performing the calculation of the triple collinear splitting functions using two independent procedures, we can also study singularities within the triple collinear splitting functions by considering their behavior in various kinematical limits. Improving our understanding of these infrared singularities is crucial in enabling the numerical evaluation of the phase space for multi-loop calculations, where explicit and implicit poles are cancelled in physical observables. We consider the iterated limit, where two partons become more collinear within the triple collinear limit, and demonstrate that the triple collinear splitting functions factorize into products of $1\to2$ splitting functions. We also consider the soft limit, where one or two partons become soft, and demonstrate how its factorization is consistent with the expectation from the soft currents~\cite{Catani:2019nqv,Catani:1999ss,Bassetto:1983mvz,DelDuca:2019ggv}. 

\subsection{Iterated Limits}

We begin by considering the iterated limit of the $1\to 3$ collinear splitting functions, where two partons become more collinear within the triple collinear limit. Without loss of generality, we assume that particles $i$ and $j$ are much more collinear to each other than either is to particle $k$. Under these conditions, the $1\to 3$ splitting functions will factorize into products of $1\to 2$ splitting functions, in general involving the polarizations of the intermediate gluon. See also, \cite{Catani:1998nv,DelDuca:2019ggv,DelDuca:2020vst,Braun-White:2022rtg,Somogyi:2005xz}. 

To impose the iterated limit within the triple collinear limit, we scale the collinear momenta of $p_i$ and $p_j$ with a new expansion parameter $\lambda^'$ by $p_{i,j}\sim p^-(\lambda^{'2},1,\lambda')$. Imposing the iterated limit by taking $\lambda' \ll \lambda$, we then extract the most singular terms from a relevant squared matrix element in Eq.~\eqref{eq:collinear_momenta}, which scales as $\sim 1 / \lambda^{\prime 2} \lambda^{2}$. Practically, this can be done by taking the triple collinear splitting functions we derived above and considering the collinear limit $\mathscr{C}_{ij}$ and identifying the most singular terms in $\lambda^'$ using the scaling of the invariants: $\tilde{s}_{ij},m_i^2,m_j^2\sim \lambda^{'2} \ll \lambda^2 \sim \tilde{s}_{ik}, \tilde{s}_{jk}, \tilde{s}_{ijk}$.

Additionally, as discussed in~\cite{Braun-White:2022rtg}, to take the iterated limit directly from the derived triple collinear splitting functions, we also find it important to take into account the azimuthal angle dependence with respect to the plane formed by the $p_i$ and $p_j$ direction. To this end, we can include the mass dependence in the azimuthal angle construction~\cite{Braun-White:2022rtg,Dulat:2018vuy} and parameterize $\tilde{s}_{ik}$ and $\tilde{s}_{jk}$ as
\begin{align}
	\tilde{s}_{ik} &\rightarrow \frac{x_i}{x_i + x_j} \tilde{s}_{123} + \frac{2 \cos \phi}{x_i - x_j} \sqrt{x_i x_j \tilde{s}_{ij} \left(\frac{x_k \tilde{s}_{a_1a_2 Q}}{1-x_k} - m^2\right)}\,, \nn \\
	 \tilde{s}_{jk} &\rightarrow \frac{x_j}{x_i + x_j} \tilde{s}_{123} + \frac{2 \cos \phi}{x_i - x_j} \sqrt{x_i x_j \tilde{s}_{ij} \left(\frac{x_k \tilde{s}_{a_1a_2 Q}}{1-x_k} - m^2\right)}\,.
\end{align}
Note that the appearance of the azimuthal angle dependence at $\mathcal{O}(\lambda^')$ is expected given that the azimuthal angle is related to the transverse momentum vectors, which are themselves $\mathcal{O}(\lambda^')$.

\subsubsection{$\bar{q} q$ Collinear Limit of $Q \rightarrow \bar{q} q Q$}
We begin by considering the $\bar{q} q$ collinear limit of the $Q \rightarrow \bar{q} q Q$ splitting function. In this case, since we factor onto a gluon intermediate state, it is essential to take spin correlations into account. The leading term in the collinear limit then takes the form
\begin{equation}
	\mathscr{C}_{q\bar{q}}\left[\frac{1}{\tilde{s}_{123}^2} {P}_{\bar{q}qQ}\right] = \frac{1}{\tilde{s}_{12} \tilde{s}_{123}} {P}_{\bar{q} q}^{\mu\nu} {H}_{g Q,\mu\nu}\,,
\label{eq:spinDepFac}
\end{equation}
where ${P}^{\mu\nu}_{\bar{q} q}$, given in \Eq{eq:SFqqSDep}, is the spin-dependent $g \rightarrow \bar{q} q$ splitting function \cite{Catani:1998nv}. The mass-dependent splitting tensor generalized from the mass-independent one given in~\cite{DelDuca:2019ggv, Somogyi:2005xz}, ${H}^{\mu\nu}_{g Q}$, is derived to be \begin{equation}
	{H}_{g Q}^{\mu \nu}(z,p,\tilde{s}) = C_F \left[\frac{1}{2} (1-z) d^{\mu \nu}(p,n) - \left(\frac{2 z}{1-z} - \frac{2 m^2}{\tilde{s}}\right) \frac{\kappa_{\perp}^{\mu} \kappa_{\perp}^{\nu}}{\kappa_{\perp}^2}\right]\,.
\end{equation}
Here, $\tilde{s}$ represents the Mandelstam variable associated with the massive `parent' quark. The symbol $d^{\mu \nu}(p,n)$ denotes the gluon polarization tensor, where $p$ is the on-shell momentum discussed below in Eq.~\eqref{eq:collinear_momenta} and $n$ is the gauge vector dependence. In general, as expected, the $n$-dependence cancels in a gauge-invariant physical quantity. The quantities $z$ and $\kappa_\perp^\mu$, respectively, signify the momentum fraction and the transverse momentum of the massive `daughter' quark from the $Q\to gQ$ splitting encapsulated by the splitting tensor. Finally, the helicity indices of $H_{ij}^{s s'}$ correspond to particle $i$ and we do not keep track of the helicity dependence of the parent heavy quark as it is trivially conserved. 

Then the azimuthal angle introduced above can be interpreted as the angle between the transverse momenta in ${P}^{\mu\nu}_{\bar{q}q}$ and ${H}^{\mu\nu}_{g Q}$, i.e. the azimuthal angle between the planes formed by $q\bar{q}$ and the one formed by $g Q$. The contraction then gives us
\begin{align}\label{eq:Qqq12CA}
	{P}_{\bar{q}q}^{\mu \nu}(x,p) {H}_{g Q, \mu \nu}(z,p',\tilde{s}_{123}) &= C_F T_R \Bigg\{\frac{1+z^2}{1-z} - \epsilon(1-z) - \frac{2 m^2}{\tilde{s}} \\
	&- 2 x(1-x) \left[1-z + 2 \left(\frac{2 z}{1-z} - \frac{2 m^2}{\tilde{s}_{123}} \right) \cos^2 \phi \right] \Bigg\}\,, \nn	
\end{align}
where $x$ and $z$ can be expressed in terms of the momentum fractions $x_i$ given in Eq.~\eqref{eq:collinear_momenta} as
\begin{equation}
	x_1 = (1-z)x,\qquad x_2 = (1-z)(1-x),\qquad x_3 = z\,.
	\label{eq:xixzRel}
\end{equation}
Performing the $d$-dimensional average over the azimuthal angle, this result can be factored into an iterated product of the spin-averaged $P_{Qg}$ and $P_{\bar{q}q}$ splitting functions. We find
\begin{equation}
\left\langle\mathscr{C}_{q\bar{q}}\left[\frac{1}{\tilde{s}_{123}^2} {P}_{\bar{q}qQ}\right]\right\rangle = \frac{1}{\tilde{s}_{12} \tilde{s}_{123}} {P}_{\bar{q}q}(x) {P}_{g Q}(1-z, \tilde{s}_{123}) = \frac{1}{\tilde{s}_{12} \tilde{s}_{123}} {P}_{\bar{q}q}(x) {P}_{Q g}(z, \tilde{s}_{123})  \,,
\end{equation}
where $\langle ~\rangle$ is used to denote the azimuthal angle averaging procedure. There are no other collinear limits within the $Q \rightarrow \bar{q}qQ$ splitting function which is as singular as the $q\bar{q}$ collinear limit.

\subsubsection{$gg$ Collinear Limit of $Q \rightarrow ggQ$}

We can apply the same procedure to understand the $gg$ collinear limit of the $Q \rightarrow ggQ$ splitting function. Here, the leading behavior is given by
\begin{align}
	\mathscr{C}_{gg}\left[\frac{1}{\tilde{s}_{123}^2} {P}_{ggQ} \right]&= \frac{1}{\tilde{s}_{12} \tilde{s}_{123}} {P}_{gg}^{\mu\nu}(x, p) {H}_{g Q, \mu \nu}(z, p', \tilde{s}_{123})\nonumber\\
	&= \frac{2 C_F C_A}{\tilde{s}_{12} \tilde{s}_{123}} \left\{\left(1 - \epsilon\right) \left(1 - z\right) \left(\frac{x}{1-x} + \frac{1-x}{x} + x\left(1-x\right)\right) \right. \nonumber \\
	&+ \left.\left(\frac{2 z}{1-z} + \frac{2 m^2}{\tilde{s}_{123}}\right) \left(\frac{x}{1-x} + \frac{1-x}{x} + 2 \left(1 - \epsilon\right) x \left(1-x\right) \cos^2 \phi\right)\right\}\,,
\end{align}
where the spin-dependent $g \rightarrow gg$ splitting function ${P}_{gg}^{\mu\nu}$ is given in~\Eq{eq:SFggSDep}. 
Averaging over the azimuthal angle and expressing the $x_i$ in terms of $x$ and $z$ as in \Eq{eq:xixzRel}, we find the factorization
\begin{equation}
\left\langle\mathscr{C}_{q\bar{q}}\left[\frac{1}{\tilde{s}_{123}^2} {P}_{ggQ}\right]\right\rangle = \frac{1}{\tilde{s}_{12} \tilde{s}_{123}} {P}_{gg}(x) {P}_{g Q}(1-z, \tilde{s}_{123}) = \frac{1}{\tilde{s}_{12} \tilde{s}_{123}} {P}_{gg}(x) {P}_{Q g}(z, \tilde{s}_{123})\,.
\end{equation}

\subsubsection{$gQ$ Collinear Limit of $Q \rightarrow ggQ$}
The triple collinear splitting function for $Q \rightarrow ggQ$ splitting also contains singular terms from the $gQ$ collinear limit. Including the spin information, when quark $i$ is collinear to the massive quark (particle $3$) the factorization takes the form
\begin{equation}
\mathscr{C}_{gQ}\left[\frac{1}{\tilde{s}_{123}^2} {P}_{ggQ}\delta^{s s'} \right]= \frac{1}{\tilde{s}_{i3} \tilde{s}_{123}} {P}_{Qg}^{h h'} {H}_{q g}^{h h'; s s'}\,.
\label{eq:QggFac}
\end{equation}
where $H_{qg}$ is given in~\cite{DelDuca:2019ggv, Somogyi:2005xz}. Here, we begin by keeping the helicity dependence of the parent heavy quark on both sides. From the collinear factorization structure of the scattering amplitude, one can show~\cite{DelDuca:2019ggv} that upon summing over the helicities of the intermediate heavy quark
\begin{align}
\delta^{h h^{\prime}} H_{qg}^{h h'; s s'}(z) = P_{qg}^{s s'}(z) = \delta^{s s'} P_{qg}(z)\,.
\end{align}
Therefore, there is no nontrivial azimuthal angle $\phi$ dependence for the $gQ$ collinear limit and the iterated limit is entirely captured by the azimuthal angle averaged case 
\begin{align}
	\left\langle \mathscr{C}_{gQ}\left[\frac{1}{\tilde{s}_{123}^2} {P}_{ggQ} \right]\right\rangle &= \frac{1}{\tilde{s}_{i3} \tilde{s}_{123}} {P}_{qg}(1 - x_j) {P}_{Qg}\left(\frac{x_3}{1-x_j}, \tilde{s}_{i3}\right) \nn \\
	&= \frac{1}{\tilde{s}_{i3} \tilde{s}_{123}} {P}_{gq}(x_j) {P}_{Qg}\left(\frac{x_3}{1-x_j}, \tilde{s}_{i3}\right) \,,
\end{align}
where we find a product of $1 \rightarrow 2$ splitting functions. As expected, the explicit computation of the $gQ$ collinear limit of the $Q \rightarrow ggQ$ splitting function shows no $\phi$ dependence. Interestingly, only the splitting function which describes the more collinear splitting is sensitive to the quark mass. Still, this provides a non-trivial check on the $1 \rightarrow 3$ splitting function mass dependence.

\subsection{Soft Limits}
In addition to iterated limits, we can also consider soft limits of the triple collinear splitting functions. We begin by considering the single gluon soft limit, where the triple collinear splitting function factorizes into a product of a $1\to 2$ splitting function and an eikonal factor. In the soft limit, the gluon $i$ has momentum scaling $p_i \sim p^- (\lambda, \lambda, \lambda)$ with an expansion parameter $\lambda$. The soft scaling then results in the following scaling
\begin{equation}
	x_i \rightarrow \lambda x_i,\;\;\;\;\; \tilde{s}_{ij} \rightarrow \lambda \tilde{s}_{ij}, \;\;\;\;\;\; p_{\perp, i}^{\mu} \rightarrow \lambda p_{\perp, i}^{\mu}\,.
\end{equation}
The leading singular term in the single soft limit then scales as $\cO(\lambda^{-2})$. Note that we do not scale the mass, as doing so makes the mass dependence subdominant.

Denoting the leading term in the limit where particle $i$ is soft by $\mathscr{S}_{i}$~\cite{DelDuca:2019ggv}, the contribution of the $Q \rightarrow ggQ$ splitting function when gluon 1 is taken soft is given by
\begin{align}
	\mathscr{S}_{g_1} & \left[\left(\frac{2 \mu^{2 \epsilon} g_s^2}{\tilde{s}_{123}}\right)^2 P_{g_1g_2Q}\right] \nn \\
	&= \left(2 \mu^{2 \epsilon} g_s^2\right) \left[\frac{2 x_3}{x_1 \tilde{s}_{13}} C_F + \left(\frac{\tilde{s}_{23}}{\tilde{s}_{12} \tilde{s}_{13}} + \frac{x_2}{x_1 \tilde{s}_{12}} - \frac{x_3}{x_1 \tilde{s}_{13}}\right) C_A \right]  \frac{2 \mu^{2 \epsilon} g_s^2}{\tilde{s}_{23}} P_{g_2Q}(x_2, \tilde{s}_{23})\,.
\end{align}
The first factor describes different eikonal contributions~\cite{DelDuca:2019ggv}. The result when gluon 2 is taken to be soft is the same as above, with $1 \leftrightarrow 2$. Note that the massive splitting function appears in the limit, which serves as a nontrivial check on the $Q \rightarrow ggQ$ mass term.

We can also consider double soft limits \cite{Catani:1999ss}, where either a quark anti-quark pair produced from an intermediate gluon or two gluons can become simultaneously soft. In general, when particles $i$ and $j$ have momenta which have soft scaling in $\lambda$, we represent the double soft limit by $\mathscr{S}_{ij}$~\cite{DelDuca:2019ggv} and the parameters involved scale as
\begin{equation}
\begin{aligned}
	x_i \rightarrow \lambda x_i, \;\;\;\;\; x_j \rightarrow \lambda x_j, \;\;\;\;\; p_{\perp, i}^{\mu} \rightarrow \lambda p_{\perp, i}^{\mu}, \;\;\;\;\; p_{\perp, j}^{\mu} 	\rightarrow \lambda p_{\perp, j}^{\mu} \\
	\tilde{s}_{ik} \rightarrow \lambda \tilde{s}_{ik}, \;\;\;\;\; \tilde{s}_{jk} \rightarrow \lambda \tilde{s}_{jk},\;\;\;\;\; \tilde{s}_{ij} \rightarrow \lambda^2 \tilde{s}_{ij}\,.
\end{aligned}
\end{equation}
The leading singular term in the double soft limit then scales as $\cO(\lambda^{-4})$. Once again, we do not scale the mass. 

For a soft $q \bar{q}$ pair, we find the following leading behavior in the limit \cite{DelDuca:2019ggv, Somogyi:2005xz}
\begin{align}
	&\mathscr{S}_{\bar{q}q}\left[\left(\frac{2 \mu^{2 \epsilon} g_s^2}{\tilde{s}_{123}}\right)^2 P_{\bar{q}qQ}\right] \\ &= \left(\mu^{2 \epsilon} g_s^2 \right)^2 \frac{8 C_F T_R}{\tilde{s}_{12}^2} \left[\frac{x_3}{x_1 + x_2} \frac{\tilde{s}_{12}}{\tilde{s}_{13} + \tilde{s}_{23}} - \frac{(x_1 \tilde{s}_{23} - x_2 \tilde{s}_{13})^2}{(x_1 + x_2)^2 (\tilde{s}_{13} + \tilde{s}_{23})^2} - m^2\frac{\tilde{s}_{12}}{(\tilde{s}_{13} + \tilde{s}_{23})^2}\right]\,, \nn
\end{align}
where the terms in the bracket contains the expected eikonal factors~\cite{DelDuca:2019ggv} computed from soft currents as well as a new mass-dependent part. For two soft gluons, the leading behavior in the limit can be separated into abelian and non-abelian contributions \cite{DelDuca:2019ggv,Catani:1999ss}
\begin{align}
	&\mathscr{S}_{gg} \left[\left(\frac{2 \mu^{2 \epsilon} g_s^2}{\tilde{s}_{123}}\right)^2 P_{ggQ}\right] \nn \\
	&= \left(\mu^{2 \epsilon} g_s^2 \right)^2 C_F \left[C_F \left( U^{gg,(\text{ab})} + V^{gg, (\text{ab})}\right)  - C_A \left(U^{gg,(\text{nab})} + V^{gg, (\text{nab})}\right)\right] \,,
\end{align}
where the mass-independent terms are in agreement with the expectations from the mass-independent part of the soft-currents and are given as
\begin{align}
U^{gg,(\text{ab})} &=\frac{16 x_3^2}{x_1 x_2 \tilde{s}_{13} \tilde{s}_{23}}\,, \\
	U^{gg,(\text{nab})} &= 8 \left(\frac{\tilde{s}_{13} \tilde{s}_{23}}{(\tilde{s}_{13} + \tilde{s}_{23})^2} + \frac{x_1 x_2}{(x_1 + x_2)^2}\right) \frac{1-\epsilon}{\tilde{s}_{12}	^2} - \frac{4 x_3}{x_2 \tilde{s}_{23}} \left(\frac{2 \tilde{s}_{23}}{\tilde{s}_{12} \tilde{s}_{13}} + \frac{2 x_2}{x_1 \tilde{s}_{12}} - \frac{2 x_3}{x_1 \tilde{s}_{13}}		\right) \nn \\ 
	&- 8 \frac{x_2 \tilde{s}_{13} + x_1 \tilde{s}_{23}}{(x_1 + x_2)(\tilde{s}_{13} + \tilde{s}_{23})} \left(\frac{1-\epsilon}{\tilde{s}_{12}^2} - \frac{x_3}{4 x_2 \tilde{s}		_{23}} \left(\frac{2 \tilde{s}_{23}}{\tilde{s}_{12} \tilde{s}_{13}} + \frac{2 x_2}{x_1 \tilde{s}_{12}} - \frac{2 x_3}{x_1 \tilde{s}_{13}}\right)\right) \nn \\
	&+ \frac{16 x_3}{(x_1 + x_2)\tilde{s}_{12}(\tilde{s}_{13} + \tilde{s}_{23})}\,,
\end{align}
and the terms with explicit mass dependence can also independently be derived from the mass-dependent part~\cite{Catani:2019nqv,Czakon:2011ve} of the soft-currents and are given as
\begin{equation}
	V^{gg, (\text{ab})} = 16 \left[- m^2 \frac{x_3}{\tilde{s}_{13} \tilde{s}_{23}} \left(\frac{1}{x_2 \tilde{s}_{13}} + \frac{1}{x_1\tilde{s}_{23}}\right) + \frac{m^4}{\tilde{s}_{13}^2 \tilde{s}_{23}^2}\right]\,,
\end{equation}
\begin{align}
	V^{gg, (\text{nab})} &= -\frac{8 m^2}{x_1 x_2 \left(x_1 + x_2 \right) \tilde{s}_{12} \tilde{s}_{13} \tilde{s}_{23} \left(\tilde{s}_{13}+\tilde{s}_{23} \right)^2} \left[x_3 (x_1+x_2)^2 \tilde{s}_{12} \left(\tilde{s}_{13} + \tilde{s}_{23}\right)\right. \nn \\
	&- \left.(x_1 + x_2) \left((x_1+x_2)^2 - 3 x_1 x_2\right)\tilde{s}_{13} \tilde{s}_{23} + \left(x_2 \left(x_1 x_2 - (x_1 + x_2)^2\right)\right)\tilde{s}_{13}^2\right.  \nn\\
	&+ \left.\left(x_1 \left(x_1 x_2 - (x_1 + x_2)^2\right)\right) \tilde{s}_{23}^2\right] + \frac{16 m^4}{\tilde{s}_{13} \tilde{s}_{23} \left(\tilde{s}_{13}+\tilde{s}_{23}\right)^2} \,.
\end{align}
Combined, these limits provide a strong check on the correctness of our results.

\section{Conclusions}\label{sec:conc}

Splitting functions are universal ingredients describing the collinear dynamics of gauge theories. As such, they play a central role in nearly all perturbative calculations in QCD, ranging from fixed order subtractions to jet substructure. While the complete set of ingredients required to describe collinear limits of massless partons are known to next-to-next-to-leading order, much less attention has been applied to the study of collinear limits of amplitudes and cross sections involving massive quarks.

In this paper, we computed the triple collinear splitting functions involving a single massive parton. Our results are compact, and ready for phenomenological applications.  Calculations were performed both by considering the collinear expansion of the relevant squared QCD matrix elements, as well as using the soft collinear effective theory with massive partons. We found exact agreement between the two methods, providing a strong cross check on our results. Additionally, we considered all iterated and soft limits of our triple collinear splitting functions, and find agreement with the predictions of factorization.

An immediate application of our results is to the improved description of jet substructure observables. While there has been tremendous recent progress in the perturbative description of the collinear substructure of jets, this has been largely restricted to the case of massless partons. Many applications, in particular Higgs and top-quark decays, as well as the study of heavy quarks in the QGP, necessitate an extension of these results to heavy flavor jets. Our calculation of the triple collinear splitting functions involving a massive parton enables the calculation of the three-point energy energy correlator on heavy flavor jets, which will be presented in a companion paper. We anticipate that this will significantly expand the scope of the precision jet substructure program, providing exciting new ways to study QCD dynamics.

\begin{acknowledgments}

We thank Stefan Hoche, Alba Soto-Ontoso, Johannes K. L. Michel, and Elliot Fox for useful discussions. We thank Prasanna Dhani, German Rodrigo and German Sborlini for communications regarding their results in \cite{Dhani:2023uxu}. K.L. was supported
by the U.S. DOE under contract number DE-SC0011090.
I.M, B.M, E.C, and M.G are supported by start-up funds from Yale University. 

\end{acknowledgments}

\appendix

\section{Additional Perturbative Ingredients}\label{app:A}

In this Appendix, we collect several additional perturbative ingredients related to the triple-collinear splitting functions, and their limits.

The massive $1\to 2$ splitting functions appear in the iterated limits of our results. They were  computed in \cite{Catani:2000ef}, but we reproduce them here for completeness. We use the notation $P_{ij}(x)$, where $x$ denotes the momentum fraction of parton $i$. The spin-averaged $1\to 2$ splitting functions are given by
\begin{align}
	{P}_{Qg}(x, \tilde{s}) &= C_F \left[\frac{1+x^2}{1-x} - \epsilon(1-x) - \frac{2 m^2}{\tilde{s}}\right]\,, \label{eq:SFQg} \\
	{P}_{gQ}(x, \tilde{s}) &= C_F \left[\frac{1+(1-x)^2}{x} - \epsilon x - \frac{2 m^2}{\tilde{s}}\right]\,, \label{eq:SFgQ} \\
	{P}_{qg}(x) &= C_F \left[\frac{1+x^2}{1-x} - \epsilon(1-x)\right]\,, \label{eq:SFqg} \\
	{P}_{gq}(x) &= C_F \left[\frac{1+(1-x)^2}{x} - \epsilon x \right]\,, \label{eq:SFgq} \\
	{P}_{\bar{q} q}(x) &= T_R \left[1 - \frac{2 x(1-x)}{1-\epsilon}\right]\,, \label{eq:SFqq} \\
	{P}_{gg}(x) &= 2 C_A \left[\frac{x}{1-x} + \frac{1-x}{x} + x(1-x)\right]\,, \label{eq:SFgg}
\end{align}
and the spin-dependent splitting functions are given by
\begin{align}
	{P}_{\bar{q} q}^{\mu \nu}(x,k) &= T_R \left[-g^{\mu \nu} + 4 x(1-x) \frac{k_{\perp}^{\mu} k_{\perp}^{\nu}}{k_{\perp}^2}\right]\,, \label{eq:SFqqSDep} \\
	{P}_{gg}^{\mu \nu}(x,k) &= 2 C_A \left[-g^{\mu \nu} \left(\frac{x}{1-x} + \frac{1-x}{x}\right) - 2 x(1-x)(1-\epsilon) \frac{k_{\perp}^{\mu} k_{\perp}^{\nu}}	{k_{\perp}	^2}\right]\,. \label{eq:SFggSDep}
\end{align}

\section{$Q \rightarrow \bar{Q}' Q' Q$ and $Q \rightarrow \bar{Q} Q Q$ Splitting Functions}\label{app:B}
In this Appendix, we also present the triple collinear splitting functions involving multiple heavy quarks as external states. Following our notation from the main text, we first consider the case where the daughter quarks $Q'$ with mass $m'$ are a different flavor than the parent quark $Q$ of mass $m$. The mass-independent piece is the same as \Eq{eq:qbqQml}, up to a prefactor
\be
	R_{\bar{Q}' Q' Q} = \frac{\tilde{s}_{12}^2}{\left(\tilde{s}_{12} + 2 {m'}^2\right)^2} R_{\bar{q} q Q}\,.
\ee
The mass dependent terms are
\begin{align}
    M_{\bar{Q}' Q' Q} &= \frac{C_F T_R}{\left(\tilde{s}_{12} + 2 {m'}^2\right)^2 \left(x_1 + x_2\right)^2} \Big\{-2 {m'}^2 \Big[\left(\tilde{s}_{13} + \tilde{s}_{23}\right) \left[\epsilon(x_1 + x_2)^3 - 2 \left(x_1(x_1^2 + 1) \right.\right. \nonumber \\
    &+ \left.\left. x_2(x_2^2 + 1) + x_1 x_2 (2 x_1 + 2 x_2 - 1)\right)\right] + 2 \tilde{s}_{13} \left(2 x_2^2 + x_1^2\right) + 2 \tilde{s}_{23} \left(2 x_1^2 + x_2^2\right) \nonumber \\
    &- \tilde{s}_{12} x_3 \left[(2 \epsilon - 3)(x_1^2 + x_2^2) + 4 (x_1 + x_2) + (4 \epsilon-2) x_1 x_2\right]\Big] - 2 m^2 \tilde{s}_{12} \left(x_1 + x_2\right)^2 \nonumber \\
    &- 4 m^2 {m'}^2 \left(x_1 + x_2\right)^2 + 4 {m'}^4 x_3 \Big[\epsilon \left(x_1 + x_2\right)^2 + 2 \left(x_1 x_2 + x_3 \left(x_1 + x_2\right)\right)\Big]\Big\}\,.
\end{align}

We also consider the case where all three quarks are the same flavor and massive. This involves another diagram which is the same as \Fig{fig:FD_qqQ} up to the exchange $(2 \leftrightarrow 3)$. The full splitting function is then given by
\be
	P_{\bar{Q}_1 Q_2 Q_3} = \left[P_{\bar{Q}'_1 Q'_2 Q_3} \Big|_{m' \rightarrow m} + \left( 2 \leftrightarrow 3 \right) \right] + P_{\bar{Q}_1 Q_2 Q_3}^{\text{(id)}}\,,
\ee
where $P_{\bar{Q} Q Q}^{\text{(id)}}$ represents the interference contributions, which are given by
\begin{align}
	R_{\bar{Q} Q Q}^{\text{(id)}} &= C_F \left(C_F - \frac{1}{2} C_A\right) \frac{\tilde{s}_{12} \tilde{s}_{13}}{\left(\tilde{s}_{12} + 2 m^2\right) \left(\tilde{s}_{13} + 2 m^2\right)} \Bigg\{ \left(1 - \epsilon\right) \left(\frac{2 \tilde{s}_{23}}{\tilde{s}_{12}} - \epsilon \right) \nonumber \\
	&+ \frac{\tilde{s}_{123}}{\tilde{s}_{12}} \left[\frac{1 + x_1^2}{1 - x_2} - \frac{2 x_2}{1 - x_3} - \epsilon \left(\frac{\left(1 - x_3\right)^2}{1 - x_2} + 1 + x_1 - \frac{2 x_2}{1 - x_3}\right) - \epsilon^2 \left(1 - x_3\right)\right] \nonumber \\
	&- \frac{x_1 \tilde{s}_{123}^2}{2 \tilde{s}_{12} \tilde{s}_{13}} \left[\frac{1 + x_1^2}{\left(1-x_2\right) \left(1-x_3\right)} - \epsilon \left(1 + \frac{2 \left(1 - x_2\right)}{1-x_3}\right) - \epsilon^2\right] \Bigg\} + \left( 2 \leftrightarrow 3 \right)\,,
\end{align}
\begin{align}
    M_{\bar{Q} Q Q}^{\text{(id)}} &= \frac{C_F \left(C_F - \frac{1}{2} C_A\right)}{\left(\tilde{s}_{12} + 2 m^2\right) \left(\tilde{s}_{13} + 2 m^2\right) \left(x_1 + x_2\right) \left(x_1 + x_3\right)} \Big\{-m^2 \Big[\tilde{s}_{12} \left[x_2 x_3 \left(2 - 2 x_2 - 3 x_3\right) \right. \nonumber \\
    &- \left. x_2 \left(4 x_2 - 5\right) - x_3 \left(x_3^2 + 4 x_3 - 5\right) - 2 + \epsilon \left(2 x_2 x_3 \left(x_2 + x_3 -2\right) + 2 x_2 \left(x_2 - 2\right) \right.\right. \nonumber \\
    &+ \left.\left. x_3\left(x_3^2 + x_3 - 3\right) + 3\right) - \epsilon^2 \left(x_1 + x_2\right)^2 \left(x_1 + x_3\right) \right] + \tilde{s}_{13} \left[x_2 x_3 \left(2 - 2 x_3 - 3 x_2\right) \right. \nonumber \\
    &- \left. x_3 \left(4 x_3 - 5\right) - x_2 \left(x_2^2 + 4 x_2 - 5\right) -2 + \epsilon \left(2 x_2 x_3 \left(x_2 + x_3 -2\right) + 2 x_3 \left(x_3 - 2\right) \right.\right. \nonumber \\
    &+ \left.\left. x_2\left(x_2^2 + x_2 - 3\right) + 3\right) - \epsilon^2 \left(x_1 + x_3\right)^2 \left(x_1 + x_2\right) \right] + \tilde{s}_{23} \left[x_2 \left(7 - x_2 - x_2^2\right) \right. \nonumber \\
    &+ \left. x_3 \left(7 - x_3 - x_3^2\right) - 4 x_2 x_3 \left(x_2 + x_3\right) - 6 + \epsilon \left(x_2^2 \left(x_2 - 2\right) + x_3^2 \left(x_3 - 2\right) \right.\right. \nonumber \\
    &+ \left.\left. 3 x_2 x_3 \left(x_2 + x_3 - 2\right) + 2\right) + \epsilon^2 \left(x_1 + x_2\right) \left(x_1 + x_3\right) \left(x_2 + x_3\right)\right] \Big] \nonumber \\
    &+ 2 m^4 \Big[x_2 \left(4 x_2 - 5\right) + x_3 \left(4 x_3 - 5\right) + x_2 x_3 \left(x_2 + x_3\right) + 2 - \epsilon\left(2 x_2 \left(x_2 - 2\right) \right. \nonumber \\
    &+ \left. 2 x_3 \left(x_3 - 2\right) - x_1 x_2 x_3 + 3\right) + \epsilon^2 \left(x_1 + x_2\right) \left(x_1 + x_3\right)\Big]\Big\} + \left(2 \leftrightarrow 3\right)\,.
\end{align}
We find agreement between our results and those of \cite{Dhani:2023uxu}.

\bibliography{1to3MassSplitFn_ref}{}
\bibliographystyle{JHEP}

\end{document}